\documentstyle[12pt]{article}
\title{Symmetries in Non Commutative Configuration space.
\thanks{Poster session at the $V^{th}$ International Conference on Mathematical Methods in Physics -IC2006, april 2006, CBPF, Rio de Janeiro and at the $XXVII^{th}$ National Meeting of Particle Physicds and Field Theory, september 2006, \'{A}guas de Lind\'{o}ia, S\^{a}o Paulo.
Also communicated by A.R.da Silva at the First Latin American Conference on Lie Groups in Geometry, june 2006, Campinas, Brazil.}}
\author{F.J. Vanhecke\thanks{vanhecke@if.ufrj.br}, C. Sigaud and A.R. da Silva \\ Instituto de F\'{\i}sica, Instituto de Matem\'{a}tica,\\ UFRJ, Rio de Janeiro, Brazil }
\def\be{\begin{equation}}
\def\ee{\end{equation}}
\def\eea{\end{eqnarray}}
\def\bea{\begin{eqnarray}}
\def\bean{\begin{eqnarray*}}
\def\eean{\end{eqnarray*}}
\def\bc{\begin{center}}
\def\ec{\end{center}}
\def\R{{\bf R}}
\def\e{{\bf e}}
\def\u{{\bf u}}
\def\v{{\bf v}}
\def\w{{\bf w}}
\def\ep{{\epsilon}}
\def\dif{{\bf d}}
\def\LG{{\cal G}}

\def\Q{{\cal Q}}
\begin{document}
\maketitle
\begin{abstract}
Extending earlier work \cite{vanhecke}, we examine the deformation of the canonical symplectic structure in a cotangent bundle $T^\star(\Q)$ by additional terms implying the Poisson non-commutativity of both configuration and momentum variables. In this short note, we claim this can be done consistently when $\Q$ is a Lie group. 
\end{abstract}
\section{Introduction}\label{eerste}
When a symplectic manifold is a cotangent bundle $\kappa:T^*(\Q)\rightarrow \Q$ with its canonical symplectic structure $\omega_0=dq^i\wedge dp_i$, the action of a diffeomorphism $\phi$ on $\Q$ induces a diffeomorphism $\Phi$ on $T^*(\Q)$ conserving $\omega_0$ :
\be\label{un}
\Phi: T^\star(\Q)\rightarrow T^\star(\Q):\{q^i,p_k\}\rightarrow \left
\{q^{\,\prime\,i}=\phi^i(q), p^\prime_k\right\}\,;\,p_l=p^\prime_k\,\frac{\partial\phi^k(q)}{\partial q^l}\ee
In particular a group action being a homomorphism $G\rightarrow {\bf Diff}(\Q)$, induces a strictly 
Hamiltonian action on $T^*(\Q)$ :
\be\label{deux}
\Phi_g:T^\star(\Q)\rightarrow T^\star(\Q):(q^i,p_k)\rightarrow \left(q^{\,\prime\,i}=\phi^i(g,q), p^\prime_k\right)\,;\,p_l=p^\prime_k\,\frac{\partial\phi^k(g,q)}{\partial q^l}\ee
Let {\bf F} be a closed two-form on configuration space, then it is well known \cite{Souriau} that a change in the symplectic structure, $\omega_0\rightarrow\omega_1=\omega_0+\kappa^\star{\bf F}$, induces a "magnetic" 
interaction without changing the "free" Hamiltonian. With this new symplectic structure, the momenta variables 
cease to Poisson commute and one needs to introduce a potential to switch to Darboux variables.\\
It is then tempting to introduce also a closed two-form in the $p$-variables in such a way that 
Poisson non commuting $q$-variables will emerge\footnote{Such an approach towards non commutative coordinates was 
originally proposed in \cite{Horvathy} in the two-dimensional case with posible application to anyon physics.}.
In this way, we obtain a (pre-)symplectic structure :
\be\label{trois}
\omega=\omega_0-\frac{1}{2}\,F_{ij}(q)\,dq^i\wedge dq^j+\frac{1}{2}\,G^{kl}(p)\,dp_k\wedge dp_l\;;\;
d\omega=0\ee
Obviously the structure of such a two-form is not maintained by general diffeomorphisms of type 
{\bf(\ref{un})}. 
But for an affine configuration space, there is the privileged group of affine transformations, $q^i\rightarrow q^{\,\prime\,i}={A^i}_j\,q^j+b^i$, which conserve such a structure. When an origin is fixed, this configuration space is identified with the translation group $\Q=G\equiv{\bf R}^N$ with commutative Lie algebra $\LG\equiv{\bf R}^N$ 
and dual $\LG^\star\equiv{\bf R}^{*N}$. Furthermore, if {\bf F} and {\bf G} are constant, $\omega$ is invariant under translations. Such a situation was examined for the N-dimensional case in our previous work \cite{vanhecke}. 
From the work of Souriau and others \cite{Souriau,Abraham,Paulette,Azcar} it is clear how to generalize the first term of this extension of the canonical symplectic two-form when configuration space is a Lie group $G$ such that phase space is trivialised $T^*G\approx G\times\LG^\star$. This is done introducing a symplectic one-cocycle, defined below.
\section{The symplectic one-cocycle}\label{tweede}
A 1-chain $\theta$ on $\LG$ with values in $\LG^\star$, on which $\LG$ acts with the  coadjoint representation {\bf k}, $\theta\in C^1(\LG,\LG^\star,{\bf k})$, is a 
linear map $\theta:\LG\rightarrow\LG^\star:\u\rightarrow\theta(\u)$.\\ 
Let $\{\e_\alpha\}$ be a basis of the Lie algebra $\LG$ with dual basis $\{\ep^\beta\}$ of $\LG^\star$ and structure constants $[\e_\alpha,\e_\beta]=\e_\mu\,^\mu{\bf f}_{\alpha\beta}$. The 1-cochain is given by  
$\theta(\u)=\theta_{\alpha,\mu}\,u^\mu\,\epsilon^\alpha$, where $\theta_{\alpha,\mu}\doteq\langle \theta(\e_\mu)|\e_\alpha\rangle$.\\
It is a 1-cocycle, $\theta\in Z^1(\LG,\LG^\star,{\bf k})$, if it has a vanishing coboundary:
\bean
(\delta_1\theta)(\u,\v)&\doteq&{\bf k}(\u)\theta(\v)-{\bf k}(\v)\theta(\u)-
\theta ([\u,\v])=0\\
\langle(\delta_1\theta)(\u,\v)|\w\rangle&\doteq&
-\,\langle\theta(\v)|[\u,\w]\rangle
+\,\langle\theta(\u)|[\v,\w]\rangle
-\,\langle\theta([\u,\v])|\w\rangle=0\\
(\delta_1\theta)_{\alpha,\mu\nu}&\doteq&\langle(\delta_1\theta)(\e_\mu,\e_\nu)|\e_\alpha\rangle\\
&\doteq&
-\,\theta_{\kappa,\nu}\;^\kappa{\bf f}_{\mu\alpha}
+\,\theta_{\kappa,\mu}\;^\kappa{\bf f}_{\nu\alpha}
-\,\theta_{\kappa,\alpha}\;^\kappa{\bf f}_{\mu\nu}=0
\eean
The 1-cocycle is called symplectic if $\Theta(\u,\v)\doteq\langle\theta(\u)|\v\rangle$ 
is antisymmetric :
\[\Theta(\u,\v)=\,-\,\Theta(\v,\u)\;;\;\Theta_{\alpha\mu}\doteq\theta_{\alpha,\mu}\]
Any antisymmetric $\Theta$ defined in terms of $\theta\in C^1(\LG,\LG^\star,{\bf k})$ is actually a 2-cochain on $\LG$ with values in {\bf R} and trivial representation : $\Theta\in C^2(\LG,{\bf R},{\bf 0})$. Furthermore, when $\theta \in Z^1(\LG,\LG^\star,{\bf k})$, $\Theta$ is a 2-cocycle of $Z^2(\LG,{\bf R},{\bf 0})$ :
\[
(\delta_2\Theta)(\u,\v,\w)\doteq\,-\,\Theta([\u,\v],\w)\,+\,\Theta([\u,\w],\v)\,-\,\Theta([\v,\w],\u)=0\]
\be\label{quatre}
(\delta_2\Theta)(\e_\alpha,\e_\beta,\e_\gamma)\doteq
\,-\,\Theta_{\kappa\gamma}\,^\kappa{\bf f}_{\alpha\beta}
\,+\,\Theta_{\kappa\beta}\,^\kappa{\bf f}_{\alpha\gamma}
\,-\,\Theta_{\kappa\alpha}\,^\kappa{\bf f}_{\beta\gamma}=0\ee
When $\LG$ is semisimple, $\Theta$ is exact. Indeed, the Whitehead lemma's state that $H^1(\LG,\R,{\bf 0})=0$ and $H^2(\LG,\R,{\bf 0})=0$. So, $\Theta$ is a coboundary of $B^2(\LG,\R,{\bf 0})$ and there exists an element $\xi$ of $C^1(\LG,\R,{\bf 0})\equiv\LG^\star$ such that 
$\Theta(\u,\v)=(\delta_1(\xi))(\u,\v)=-\,\xi([\u,\v])$ or 
$\Theta_{\alpha\beta}=-\,\xi_\mu\;^\mu{\bf f}_{\alpha\beta}$.\\ 
In general, $\Theta=\frac{1}{2}\,\Theta_{\alpha\beta}\,\ep^\alpha\wedge\ep^\beta$, with $\Theta$ obeying the cocycle condition {\bf(\ref{quatre})}. Acting with 
${L^\star}_{g^{-1}|g}:T^\star_{e}(G)\rightarrow T^\star_g(G)$, yields the left-invariant forms :
\bean
\ep^\alpha_L(g)&\doteq&{L^\star}_{g^{-1}|g}\,\ep^\alpha={L^\alpha}_\beta(g^{-1};g)\,\dif g^\beta\\
\Theta_L(g)&\doteq&{L^\star}_{g^{-1}|g}\;\Theta=
(1/2)\,\Theta_{\alpha\beta}\;\ep^\alpha_L(g)\wedge\ep^\beta_L(g)\eean
where $\;{L^\alpha}_\beta(g;h)\doteq\partial(gh)^\alpha/\partial h^\beta$. 
Using the cocycle relation {\bf(\ref{quatre})} and the Maurer-Cartan structure equations,   
\[\dif\ep^\alpha_L(g)=-\,\frac{1}{2}\;^\alpha{\bf f}_{\mu\nu}\,\ep^\mu_L(g)\wedge
\ep^\nu_L(g)\] it is seen that $\Theta_L(g)$ is a closed left-invariant two-form on $G$. 
\section{G Actions on $T^\star(G)$}\label{derde}
Natural coordinates of points $x=(g,{\bf p})\in T^\star(G)$ are given by 
$(g^\alpha,p_\beta)$, where ${\bf p}=p_\beta\,\dif g^\beta$. There are two canonical trivialisations of the cotangent bundle. 
\begin{itemize}
\item
The left trivialisation : 
\[\lambda:T^\star(G)\rightarrow G\times\LG^\star:
(g,p_g)\rightarrow\left(g,\pi^L={L^\star}_{g|e}\;p_g={\pi^L}_\mu\,\ep^\mu\right)_{\bf B}\]
which yields "body" coordinates, given by $(g^\alpha,{\pi^L}_\mu)_{\bf B}$.
\item
The right trivialisation : 
\[\rho:T^\star(G)\rightarrow G\times\LG^\star:
(g,p_g)\rightarrow\left(g,\pi^R={R^\star}_{g|e}\;p_g={\pi^R}_\mu\,\ep^\mu\right)_{\bf S}\]
which yields "space" coordinates, given by $(g^\alpha,{\pi^R}_\mu)_{\bf B}$.
\end{itemize}
They are related by : 
$\pi^R={R^\star}_{g^{-1}|g}\circ {L^\star}_{g|e}\;\pi^L={\bf K}(g)\;\pi^L$, 
where ${\bf K}(g)$ is the coadjoint representation of $G$ in $\LG^\star$.\\
Lifting the left multiplication of $G$ by $G$ to the cotangent bundle yields 
\[\Phi^L_a:T^\star(G)\rightarrow T^\star(G):
x=(g,p_g)\rightarrow y=(ag,p^{\,\prime}_{ag}=L^\star_{a^{-1}|ag}\,p_g)\]
From $\lambda\circ L^\star_{a^{-1}|ag}:p_g\rightarrow L^\star_{ag|e}\circ L^\star_{a^{-1}|ag}\;p_g=L^\star_{g|e}\;p_g=\pi$, it is seen that, in body coordinates, 
$\left(\Phi^L_a\right)_{\bf B}\doteq\lambda\circ\Phi^L_a\circ\lambda^{-1}:(g,\pi^L)_{\bf B}\rightarrow (ag,\pi^L)_{\bf B}$. \\
The pull-back of the cotangent projection $\kappa:T^\star(G)\rightarrow G:x\doteq(g,{\bf p})\rightarrow g$, yields differential forms on the cotangent bundle : 
\bea
\langle\ep^\alpha_L(x)|&=&\kappa_x^\star\;\ep^\alpha_L(\kappa(x))\nonumber\\
\label{cinq}
\widetilde{\Theta}_L(x)&=&\kappa_x^\star\;\Theta_L(\kappa(x))=
-\;\frac{1}{2}\;\Theta_{\alpha\beta}\;
\langle\ep^\alpha_L(x)|\wedge\langle\ep^\beta_L(x)|\eea 
Since $\Theta(g)$ is closed on $G$, its pull-back, $\widetilde{\Theta}_L(x)$, is 
a closed 2-form on $T^\star(G)$.\\
Furthermore, the left-invariance of 
$\ep^\alpha_L(g)$ : $L^\star_{a^{-1}|ag}\;\ep^\alpha(g)=\ep^\alpha(ag)$ implies the 
$\Phi^L_a$-invariance of its pull-back : 
$(\Phi^L_a)^\star_{|x}\;\langle\ep^\alpha_L(\Phi^L_a(x))|=\langle\ep^\alpha_L(x)|$ and so is 
$\widetilde{\Theta}_L(x)$. A $\Phi^L_a$-invariant basis 
of one-forms on $T^\star(T^\star(G))$ is 
\be\label{six}\{\langle\ep^\alpha_L|;\langle\dif{\pi^L}_\mu|\}\ee
The right multiplication by $a^{-1}$ induces another {\it left} action by :
\[\Phi^R_a:T^\star_g(G)\rightarrow T^\star_{ga^{-1}}(G):
(g,p_g)\rightarrow(ga^{-1},p^{\,\prime}_{ga^{-1}}=R^\star_{a|ga^{-1}}\,p_g)\;,\]
Computing : $L^\star_{ga^{-1}|e}\circ R^\star_{a|ga^{-1}}\circ L^\star_{g|e}\;\pi^L=
L^\star_{a^{-1}|e}\circ R^\star_{a|a^{-1}}\;\pi^L$, it follows that, in body coordinates,
$\Phi^R_a$ acts as : $\Phi^R_a:(g,\pi^L)_{\bf B}\rightarrow(g^\prime=ga^{-1},{\pi^\prime}^L=
{\bf K}(a)\pi^L)_{\bf B}$.
Under $\Phi^R_a$, the $\Phi^L_a$-invariant basis {\bf(\ref{six})} transforms as 
\bea\label{sept}
(\Phi^R_a)^\star_{|x}\;\langle\ep^\alpha_L(\Phi^R_a(x))|&=&{{\bf Ad}^\alpha}_\beta(a)\;
\langle\ep^\beta_L(x)|\nonumber\\
(\Phi^R_a)^\star_{|x}\;\langle \dif{\pi^{\prime L}}_\mu|&=&\langle\dif{\pi^L}_\nu|\;
{{\bf Ad}^\nu}_\mu(a^{-1})\eea
\section{The modified symplectic structure on $T^\star(G)$}\label{vierde}
The canonical Liouville one-form on $T^\star(G)$ and its associated  symplectic two-form are   
$\;\langle\theta_0|=p_\alpha\;\langle dg^\alpha|=\pi_\mu\;\langle\ep^\mu_L|$, and
\bea\label{huit}
\omega_0&=&-\,\dif\langle\theta_0|=
-\pi_\mu\,\dif\langle\ep^\mu_L|+
\langle\ep^\mu|\wedge\langle\dif\pi_\mu|\nonumber\\
&=&
\frac{1}{2}\,\pi_\mu\;^\mu{\bf f}_{\alpha\beta}\;
\langle\ep^\alpha|\wedge\langle\ep^\beta|
+
\langle\ep^\mu|\wedge\langle\dif\pi_\mu|\eea
A modified symplectic two-form is obtained adding the closed two-form {\bf(\ref{cinq})}, constructed from the symplectic cocycle:
\be\label{neuf}
\omega=\omega_0+\widetilde{\Theta}_L=
\frac{1}{2}\,\left(\pi_\mu\;^\mu{\bf f}_{\alpha\beta}+\Theta_{\alpha\beta}\right)\;
\langle\ep^\alpha|\wedge\langle\ep^\beta|
+
\langle\ep^\mu|\wedge\langle\dif\pi_\mu|\ee
For semisimple $\LG$, this reduces to :
\be\label{dix}
\omega=
\frac{1}{2}\,(\pi_\mu-\xi_\mu)\;^\mu{\bf f}_{\alpha\beta}\;
\langle\ep^\alpha|\wedge\langle\ep^\beta|
+
\langle\ep^\mu|\wedge\langle\dif\pi_\mu|=
-\,\dif\,\left((\pi_\mu-\xi_\mu)\;\langle\ep^\mu_L|\right)
\ee
This means that the Liouville form is modified 
$\langle\theta_L|=\left((\pi_\mu-\xi_\mu)\;\langle\ep^\mu_L|\right)$ such that $\omega=\,-\,\dif\langle\theta_L\,|$ and that $\{g,p^\prime_g\doteq{L^\star}_{g^{-1}|g}(\pi-\xi)\}$ and there are global Darboux coordinates : $\{g^\alpha,{p^\prime}_\mu=p_\mu-\xi_\beta\,{L^\beta}_\mu(g^{-1};g)\}$.\\
Finally we may add another left-invariant and closed two-form in the $\pi$ variables $\widetilde{\Upsilon}_L=(1/2)\,\Upsilon^{\mu\nu}\,\langle\dif\pi_\mu|\wedge\langle\dif\pi_\nu|$ such that 
\be\label{onze}
\omega_L=\omega_0+\widetilde{\Theta}_L+\widetilde{\Upsilon}_L\ee
defines a $\Phi^L_a$-invariant (pre)-symplectic two form on $T^\star(G)$.\\
Under $\Phi^R_a$, this (pre-)symplectic two-form {\bf(\ref{douze})} is invariant if $a$ belongs to the intersection of the isotropy groups of 
$\widetilde{\Theta}_L$ and $\widetilde{\Upsilon}_L$ :
\be\label{douze}
\Theta_{\alpha\beta}\;{{\bf Ad}^\alpha}_\mu(a)\,{{\bf Ad}^\beta}_\nu(a)=
\Theta_{\mu\nu}\;;\;
{{\bf Ad}^\alpha}_\mu(a^{-1})\,{{\bf Ad}^\beta}_\nu(a^{-1})\;\Upsilon^{\mu\nu}=
\Upsilon^{\alpha\beta}\ee
\section{Conclusions}\label{vijfde}
The degeneracy of the two-form {\bf(\ref{onze})} will be examined in further work, as was done in \cite{vanhecke} for the abelian group. If $\omega_L$ is not degenerate, Poisson Brackets can be defined and, in the degenerate case, the constrained formalism of \cite{GNH} is applicable. Finally, if the isotropy group of {\bf(\ref{douze})} is not empty, the remaining $\Phi^R_a$-invariance will provide momentum mappings. Equations of motion of the Euler type will follow from a Hamiltonian of the form 
\[H\doteq\frac{1}{2}\,{\cal I}^{\mu\nu}\;{\pi^L}_\mu\,{\pi^L}_\nu\]
The momenta mentionned above will be conserved if the isotropy group above also conserves the {\it inertia tensor} ${\cal I}$.  

\end{document}